\documentclass[prb,twocolumn,amsmath,showpacs]{revtex4}

\usepackage{graphicx}

\begin{document}
\input epsf

\title{
The magnetic flux dynamics in critical state of one-dimensional
discrete superconductor}

\author {S. L. Ginzburg, A. V. Nakin, N. E. Savitskaya}
\affiliation{Petersburg Nuclear Physics Institute, 188300,
Gatchina, Russia}

\date{\today}

\begin{abstract}

We give a theoretical description of avalanche-like dynamics of magnetic
flux  in the critical state of ``hard" type-II superconductors using a
model of a one-dimensional multijunction SQUID that well reproduces
 the main magnetic properties of these objects. We show that the system
under consideration demonstrates the self-organized criticality. The
avalanches of vortices manifest themselves as jumps of the total
magnetic flux in the sample. The sizes of these jumps have a power-law
distribution. Our results are in qualitative agreement with
experiments.

\end{abstract}
\pacs{05.65.+b, 74.50.+r}

\maketitle

\section{Introduction}

About three decades ago  Charles P. Bean proposed a model describing a
critical state in  ``hard" type-II superconductors.\cite{Bean} In the
frame of this model, the critical state, arising in a superconductor
as a result of interplay of magnetic and pinning forces, is a
non-equilibrium state, in which the density of Abrikosov vortices
varies linearly with the distance from the sample surface. Later Pierre
G.~de~Gennes considered an interesting analogy between the critical
behaviour of magnetic flux in superconductors and evolution of a sand
pile.\cite{DeGennes} De~Gennes compared the magnetic flux dynamics in
the critical state with the process that is observed in an
over-critical sand pile where sand slides from the top of the pile
forming avalanches. According to this brilliant concept, when the
superconductor becomes over-critical, the vortices begin to move in
form of avalanches. Each avalanche includes a simultaneous motion of a
large number of vortices.

Although this picture of the magnetic dynamics in superconductors was
formulated in 60-th, the  experimental technique, allowing observation
of vortex avalanches and studying  their size distribution, was
developed only in the last decade (see, for example, Ref.
\onlinecite{Alt}). These experiments demonstrated that in certain
conditions slow changes of an applied magnetic field result in an
avalanche-like redistribution of vortices. It turns out that the sizes
of avalanches have a power-law distribution.\cite{Feld, Aeger, Matizen}
Such a distribution is, as is well known, a characteristic feature of
systems with a self-organized criticality (SOC).\cite{Bak}

The concept of SOC was formulated in 1987 in order to  explain a
behavior of giant dissipative dynamical systems consisting of a large
number of interacting elements.\cite{Bak}  According to the main
principles of this concept, such systems are able to accumulate small
external perturbations and evolve into a self-reproducing  critical
state. This state is an ensemble of various metastable states. The
critical system migrates from one metastable state to the other by
means of avalanches. The avalanches are initiated by small external
perturbations and may have various sizes. The distribution of
avalanche sizes is  a power-law function with an exponent close to 1.
The paradigm of SOC is a sandpile.  This is why the corresponding
mathematical model is called the sandpile model.\cite{Bak,Dhar}

Recently, considering the similarity between the critical magnetic
dynamics in superconductors and SOC, we showed that  employing of
dynamical equations provides a promising way to correlate these two
phenomena. \cite{Ginzburg,JETF,JLTP} In these  works, we used the fact
that the main features in the behavior of a ``hard" type-II
superconductor may well be reproduced by considering a so called
discrete superconductor, i.e., a lattice of Josephson junctions
(multijunction SQUID).\cite{Chen,Ginzburg}  An additional advantage of
this approach is that the corresponding equations are not difficult
for the analysis.

It is known that magnetic properties of a lattice of Josephson
junctions may be characterized by a parameter $V\sim j_ca^3/\Phi_0$
($j_c$ is the critical current density of the junction, $\Phi_0$
is the magnetic flux quantum, and $a$ is the interjunction
distance).\cite{Chen,Wolf} In the case of $V\ll1$, the system can be
considered as a  single sandwich-like Josephson junction without
pinning.\cite{Chen1}  If the condition $V\ll1$ is not satisfied, a
multijunction SQUID, due to its discreteness,  is able to pin magnetic
flux.  In the case of $V\gg1$, we have a system with strong pinning.

It was also  shown theoretically and by computer simulations that the
critical state of two-dimensional (2D) and one-dimensional (1D)
multijunction SQUID's with $V\gg1$ is
self-organized.\cite{Ginzburg,JETF,JLTP} Avalanches in such systems
manifest themselves as pulses of voltage across the
junctions.\cite{Ginzburg, JETF}  It was demonstrated that the total
voltage integrated over the time of the avalanche is an analog of the
avalanche size in the sandpile model.\cite{Ginzburg} For some way of
perturbation, this quantity has a power-law distribution. It was also
proven that for $V\gg1$ the dynamical equations for the
superconductor are equivalent to algorithms describing the sandpile
model.

The quantities usually  measured in the  sandpile experiment are the
mass of the pile and its variations. \cite{Held} The direct analog for
the multijunction SQUID is the total magnetic flux $\Phi$  in the
sample and its fluctuations.  This characteristic  was studied
experimentally, for instance, in Refs. \onlinecite{Feld} and
\onlinecite{Matizen}. This is why, in this work, we concentrate on
calculation of $\Phi$.

The aim of  this work is a theoretical investigation and a computer
simulation of the magnetic flux dynamics and the avalanche statistics
in the critical state of a multijunction SQUID. We use a model of a 1D
multijunction SQUID with intrinsic spatial randomness, which was
introduced  in Ref. \onlinecite{JLTP}. This model takes into account
that  real superconductors are disordered systems. A spatial
randomness, introduced in our model, is an equivalent of the existence
of randomly distributed pinning centers.\cite{JLTP}.  We consider here
not only the case of $V\gg1$, studied
earlier,\cite{Ginzburg,JETF,JLTP} but also $V\lesssim1$.  We show
that for all values of $V$ and for a fixed degree of disorder, the
critical state of the system may be represented as a set of metastable
states transforming to each other due to avalanches. An avalanche in
such a system results in penetration of magnetic flux into
superconducting sample and its redistribution between the system
cells. We demonstrate that for all considered values of $V$ the
probability density of magnetic flux jumps has a power-law
distribution. Thus, using the simplest model of a superconductor, we
obtain the results that are in qualitative agreements with
experiments.

 The paper is organized as follows. The second section is devoted to
description of our model of 1D multijunction SQUID with intrinsic
spatial randomness.  We analyze the dynamical equations and
demonstrate that disorder of the system  is a key factor for
realization of the complex critical state.  In the third section, we
discuss the structure of the critical state for various values of the
SQUID-parameter $V$. Single avalanches are described in the forth
section.  In the fifth section, we analyze the statistics of magnetic
flux jumps and discuss the possibility of the realization of the
self-organized critical state in the system under consideration.  The
results of calculations are compared with experiments. In Conclusions
we summarize the main results of the work.

\section{Basic equations}

\begin{figure}
\centerline{\epsfxsize=8cm \epsfysize=6cm\epsfbox{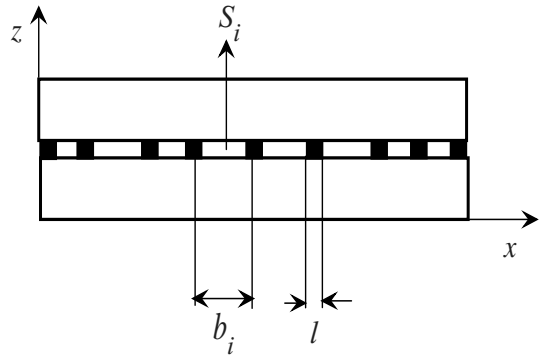}}
\caption{The $x-z$ section of 1D multijunction SQUID
\label{fig:fig1}}
\end{figure}

The one-dimensional multijunction SQUID, which we consider here, may be
represented as two infinitely long in the $y$-direction superconducting
 layers connected by Josephson junctions, as is shown in Fig.
\ref{fig:fig1}. All junctions have the same length $l$ along the
$x$-axis.  The distance between the $i$-th and the $i+1$-th junctions
we denote as a random variable $b_i$. The system is placed in a
slowly increasing external magnetic field $H_{\rm ext}$, aligned along the
$y$-axis.  Using the resistive model of a Josephson junction and
neglecting by thermal fluctuations, we can write the current density
$j_i$ as:
\begin{equation} \label{j}
j_i=j_c\sin\varphi_i+\frac{\Phi_0}{2\pi\rho}
\frac{\partial\varphi_i}{\partial t},
\end{equation}
where $j_c$ is the critical current density, $\varphi_{i}$ is the
gauge-invariant phase difference across the $i$-th junction, $\rho$ is
the junction resistance per unit area. The  current density $j_i$ is
connected with the magnetic field $H_i$ in the  neighboring cells by
the following expression (we numerate the cell by the nearest left
junction):
\begin{equation} \label{F}
4\pi j_i=\frac{H_i-H_{i-1}}{l}=
\left(\frac{\Phi_i}{S_i}-\frac{\Phi_{i-1}}{S_{i-1}
}\right)\frac1l,
\end{equation}
 where $\Phi_{i}=H_{i}S_{i}$ is the  magnetic flux in the  $i$-th cell,
$S_{i}=2\lambda b_{i}$ is the cell area, $\lambda$ is the magnetic
field penetration depth.\cite{JLTP}

Taking into account that
\begin{equation}
\Phi_i(t)=\frac{\Phi_0}{2\pi}\left[\varphi_{i+1}(t)-\varphi_i(t)\right],
\label{pot}
\end{equation}
we obtain  the system of differential equations for the gauge-invariant
phase differences:
\begin{eqnarray}
&& V\sin\varphi_i+\tau\frac{\partial\varphi_{i}}{\partial t}\quad =
\nonumber\\
&&\left[J_i\varphi_{i+1}-(J_i+J_{i-1})\varphi_i+J_{i-1}\varphi_{i-1}
\right];\ i\neq 1,N;
\nonumber\\
&& V\sin\varphi_1+\tau\frac{\partial\varphi_{1}}{\partial t}=
[J_{1}\varphi_2-J_1\varphi_{1}]-2\pi h_{\rm ext};
\label{main}\\
&&\hspace*{-0.4cm} V\sin\varphi_N+\tau\frac{\partial\varphi_N}{\partial
t}= [-J_{N-1}\varphi_N+J_{N-1}\varphi_{N-1}]+2\pi h_{\rm ext};
\nonumber\\
&& V=\frac{16\pi^2 al\lambda j_c}{\Phi_0};\quad
\tau=\frac{8\pi al\lambda}\rho;
\nonumber\\
&& J_i=\frac a{b_i}; \quad h_{\rm ext}=\frac{2\lambda a}{\Phi_0}H_{\rm ext},
\nonumber
\end{eqnarray}
where $a=\langle b_i\rangle$ is the average interjunction distance.

In order to take into account a spatial disorder of real physical
systems, we consider the distances between the junctions as random.
As may be seen from Eqs. (\ref{main}), in our model this randomness is
equivalent to a scatter of the coefficients $J_{i}$. In the following,
we assume that the values of these coefficients are distributed
uniformly in the interval $[1,1+\Delta J]$.

\begin{figure}
\centerline{\epsfxsize=6cm \epsfysize=8cm\epsfbox{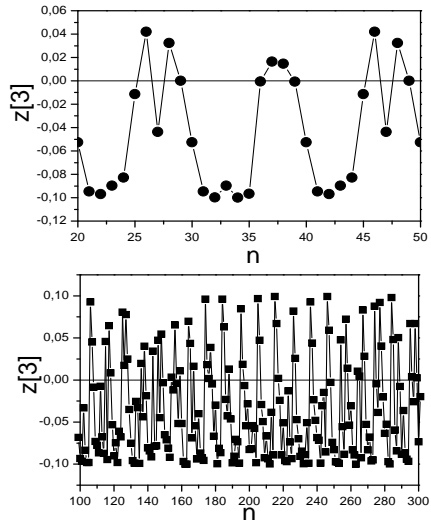}}
\caption{The dependence of the dimensionless current $z[k]$ for $k =
3$ and $V=0.6$ on the number of avalanches $n$ for $\Delta J = 0$
(upper panel) and $\Delta J=0.1$ (lower panel) scattering of the
coefficients $J_{i}$. We see that introducing of disorder results in a
crucial increase of the number of metastable states.
\label{fig:fig7}}
\end{figure}

As was stated above, there must be a large number of metastable states in
the system in order to expect a  self-organized behavior. In our case,
such a situation can be realized only if $\Delta J_i\neq0$. This is
illustrated in Fig.~\ref{fig:fig7},  which shows the dimensionless
current $z[k]=V/2\pi\sin\varphi_k$ in one of the junctions at the
final moment of $n$-th avalanche for two cases:  (i)~for a regular
system with $\Delta  J_{i}=0$ and (ii)~for a disordered system with
$\Delta J=0.1$.  We see that, while the regular system behaves itself
periodically and has only a limited number of possible  values of the
junction current, the situation for $\Delta J\neq0$ is completely
different.  In the latter case, the number of possible values of
$z_k$ is rather large.  This result does not depend, of course, on
the number of the chosen junction.

\section{Structure of the critical state of 1D multijunction SQUID}

In this section, we present the result of computer simulation of the
critical state for the system described above. Four different values
of the SQUID-parameter are considered. We take $V=40$ in order to
model the case of $V\gg 1$ and $V=0.3$, $V=0.6$, and $V=1.2$ to
simulate the situation with $V \lesssim1$.

We  use an Euler integration  scheme for Eqs. (\ref{main}) with $\Delta
t=0.01$, for $V\gg1$ and with $\Delta t=0.1$ for  $V\lesssim1$.
The non-stationary method of perturbation, which is commonly used for
simulations of self-organized systems, is employed.  It means that the
external magnetic field $h_{\rm ext}$ varies only when all relaxation
processes in the system are already finished. This way of changing  of
the external field is close to that in experiments of Aegerter et
al.\cite{Aeger}

Before starting,   we fix a set of random coefficients  $J_{i}$, which
remain unchanged during the simulation process.  Starting from the
state with $\varphi_{i}=0$, we perturb the system by increasing the
external field from $h_{\rm ext}$ to $h_{\rm ext}+\Delta h_{\rm ext}$. In our
simulations, we use $\Delta h_{\rm ext}=1$ for $V=40$ and $\Delta
h_{\rm ext}=0.1$ for small and transient values of $V$.  Then we allow the
system to relax and, as was mentioned above, we assume that the value
of $h_{\rm ext}$ is constant during the relaxation process.  We take that
the system has already reached its metastable state if
$d\varphi_i/dt<10^{-7}$ for all $i$.   When the dynamics stops we
perturb the system again and so on.

First, we analyze  the distribution of magnetic field  inside our SQUID
for various values of $V$. The magnitude of the dimensionless magnetic
field in the $i$-th cell,  measured at the end of $n$-th avalanche,
may be calculated as:
\begin{equation}
h_i^{(n)}\ =\ \frac{2\lambda a\Phi_i^{(n)}}{S_i\Phi_0}\ .
\end{equation}
 Note that for all values of $V$ and $\Delta J$ the system demonstrates
irreversible magnetic behavior. Remanent magnetization becomes zero
only for negligible value of the parameter $V$ ($V=0.06$) and for
$\Delta J_{i}=0$.

\begin{figure}
\centerline{\epsfxsize=6cm\epsfysize=10cm\epsfbox{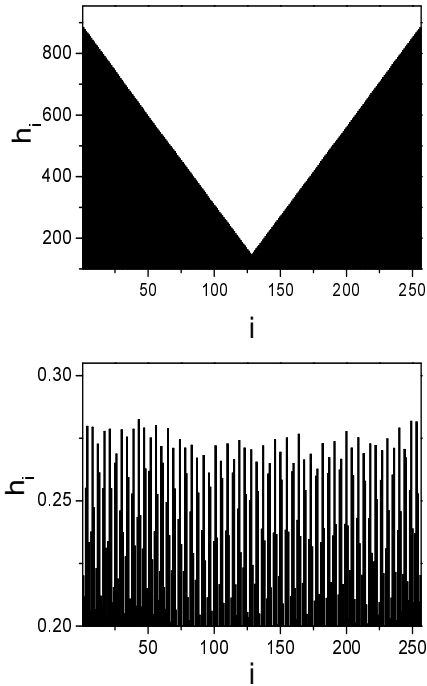}}
\caption{The magnetic field profiles
$h_i^{(n)}$ inside the SQUID for $V=40$ (upper panel) and
$V=0.3$ (lower panel), $N=257$.
\label{fig:fig3}}
\end{figure}

 Fig. \ref{fig:fig3} shows the profiles of magnetic field inside the
SQUID for one of metastable states for two different values of $V$ and
$\Delta J=0.1$.  We see that for $V=40$, when every cell acts as
pinning center, the profile is similar to the result of Bean's model.
In the case of $V=0.3$, the pinning centers are represented by groups
of cells and a number of peaks may be seen. We note that the peak
amplitudes are different for different metastable states.

Now we consider  the dynamics of the total magnetic flux in the sample
and its dependence on the external magnetic field $h_{\rm ext}$.
The total magnetic flux may be calculated as
\begin{equation}
\Phi^{(n)}=\sum_{i=1}^{N-1}\Phi_{i}^{(n)}=\frac{\Phi_{0}}{2\pi}
\left(\varphi_{N}^{(n)}-\varphi_{1}^{(n)}\right).
\end{equation}

\begin{figure}
\centerline{\epsfxsize=8cm\epsfysize=10cm\epsfbox{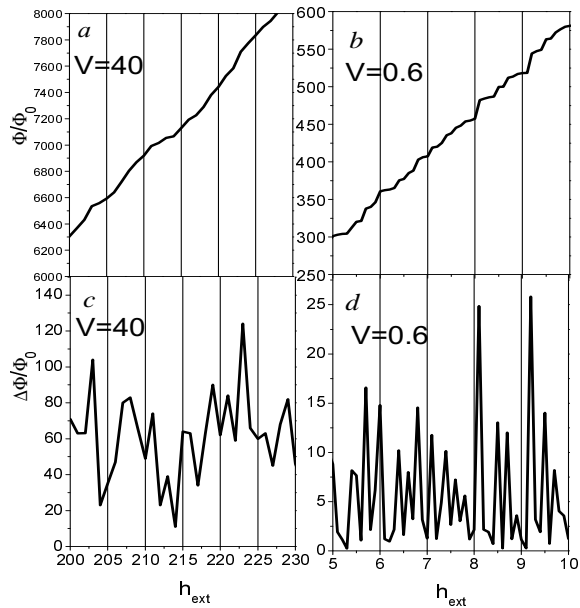}}
\caption{The total magnetic flux in the SQUID (a--b) and the magnitude
of the flux jumps (c--d) as functions
of the external magnetic field.  The SQUID size $N=65$ and $\Delta
J=0.1$.  \label{fig:potok}}
\end{figure}

Here $\varphi_{i}^{(n)}$ denotes the value of the phase for the final
moment of $n$-th avalanche. The variation of the total magnetic flux
due to  the $n$ avalanche
\begin{equation}
\Delta\Phi^{(n)}\ =\ \Phi^{(n)}-\Phi^{(n - 1)}.
\end{equation}
Figures \ref{fig:potok}(a) and \ref{fig:potok}(b) illustrate the
evolution of the total magnetic flux with increasing magnetic field
for two systems with different values of $V$.  As was noted earlier,
the external magnetic field is assumed to be constant during the
avalanche. The values, plotted in Fig. \ref{fig:potok}, correspond to
the moments when the relaxation is already finished but a new step of
the external field is not yet applied. Fig. \ref{fig:potok}(c) and
\ref{fig:potok}(d) show the corresponding dependencies of $\Delta
\Phi$. As may be seen, there are avalanches of different sizes.  We
also see that the behavior remains qualitatively the same for different
values of $V$.

\begin{figure}
\centerline{\epsfxsize=6cm \epsfysize=6cm\epsfbox{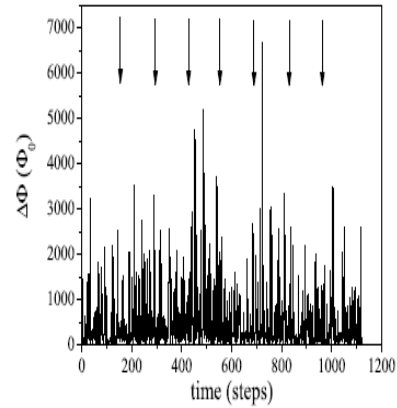}}
\caption{$\Delta\Phi$ as a function of time from Ref.
\onlinecite{Aeger}. The data presented in this figure come from series
of nine experiments and different experimental runs are separated by
vertical arrows. \label{fig:aeger}}
\end{figure}

In Fig. \ref{fig:aeger}, we show experimental results of
Ref. \onlinecite{Aeger}.  In this work, the dynamics of the magnetic
flux in the critical state of a thin ${\rm YBa_2Cu_3O}_{7-x}$ film
was studied.  The external magnetic field was changed in step-wise
manner.  After each field step of 0.5\,Oe the sample was allowed to
relax for 10 seconds before the magnetic flux was measured. These
experimental conditions are similar to rules that we use in our
computer simulation.  As may be seen in Fig.~\ref{fig:aeger}, similar
to that in computer simulation, experimentally measured magnitudes of
the magnetic flux jumps have various sizes.

\begin{figure}
\centerline{\epsfxsize=8cm \epsfysize=6cm \epsfbox{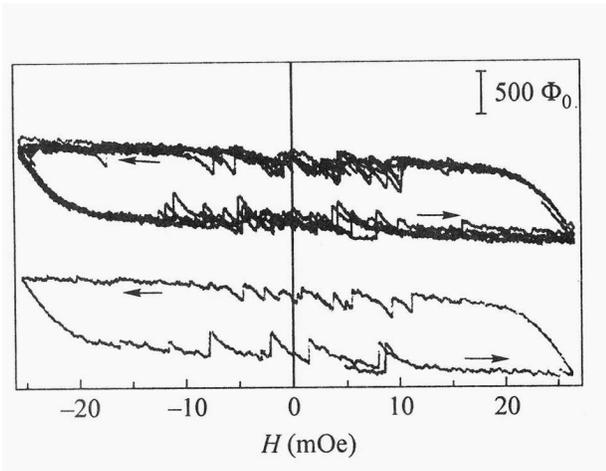}}
\caption{The histeresis loops of the total magnetic moment in the 2D
lattice of Josephson junctions from Ref. \onlinecite{Matizen}.
\label{fig:matizen}}
\end{figure}

The magnetic flux jumps was also observed in an artificial 2D lattice
of Josephson junctions.\cite{Matizen}   In these experiments, the
external magnetic field was changed continuously but slowly. This
allows to neglect by the change of the magnetic field during the
relaxation process. Fig.~\ref{fig:matizen} shows the hysteresis loops
of the total magnetic moment of the lattice from Ref.
\onlinecite{Matizen}. It may be seen that the avalanches, which are
represented by jumps of the magnetic moment, have different sizes and
include hundreds of magnetic flux quanta.  The upper trace shows
several superposed hysteresis loops for the same experimental
conditions. Due to this superposition, we can see the randomness in
jumps. The random jumps with different sizes may be considered as a
manifestation of SOC. As may be seen comparing Figs.~\ref{fig:potok}
and \ref{fig:matizen}, the results of our simulation are in qualitative
agreement with experiments.

\section{The avalanche structure in the critical state}

As was demonstrated above, an increase of the external magnetic field can
launch an avalanche. In this way  the system migrates form one metastable
state to the other. An avalanche in our multijunction SQUID is a
simultaneous penetration of a considerable number  of vortices in the
sample and a redistribution of the corresponding magnetic flux inside
the system. As a result of such a process, the system reaches a next
metastable state, the total magnetic flux increases and the values of
the magnetic flux in some cells change. In this section, we consider
the process of the avalanche development.

At every moment $t_n$ during the $n$-th avalanche we calculate
the magnetic flux $\Phi_i^{(n)}(t_n)$ using the  expression
(\ref{pot}). We also calculate the difference
$\Delta\Phi_i^{(n)}(t_n)=[\Phi_i^{(n)}(t_n)-\Phi_{i}^{(n)}(t_{0n})]$
where $t_{0n}$ is the initial moment of the avalanche. We emphasize that
all the results in this section are related to a single avalanche and the
external magnetic field is assumed to be constant.

\begin{figure}
\centerline{\epsfxsize=8cm \epsfysize=8cm\epsfbox{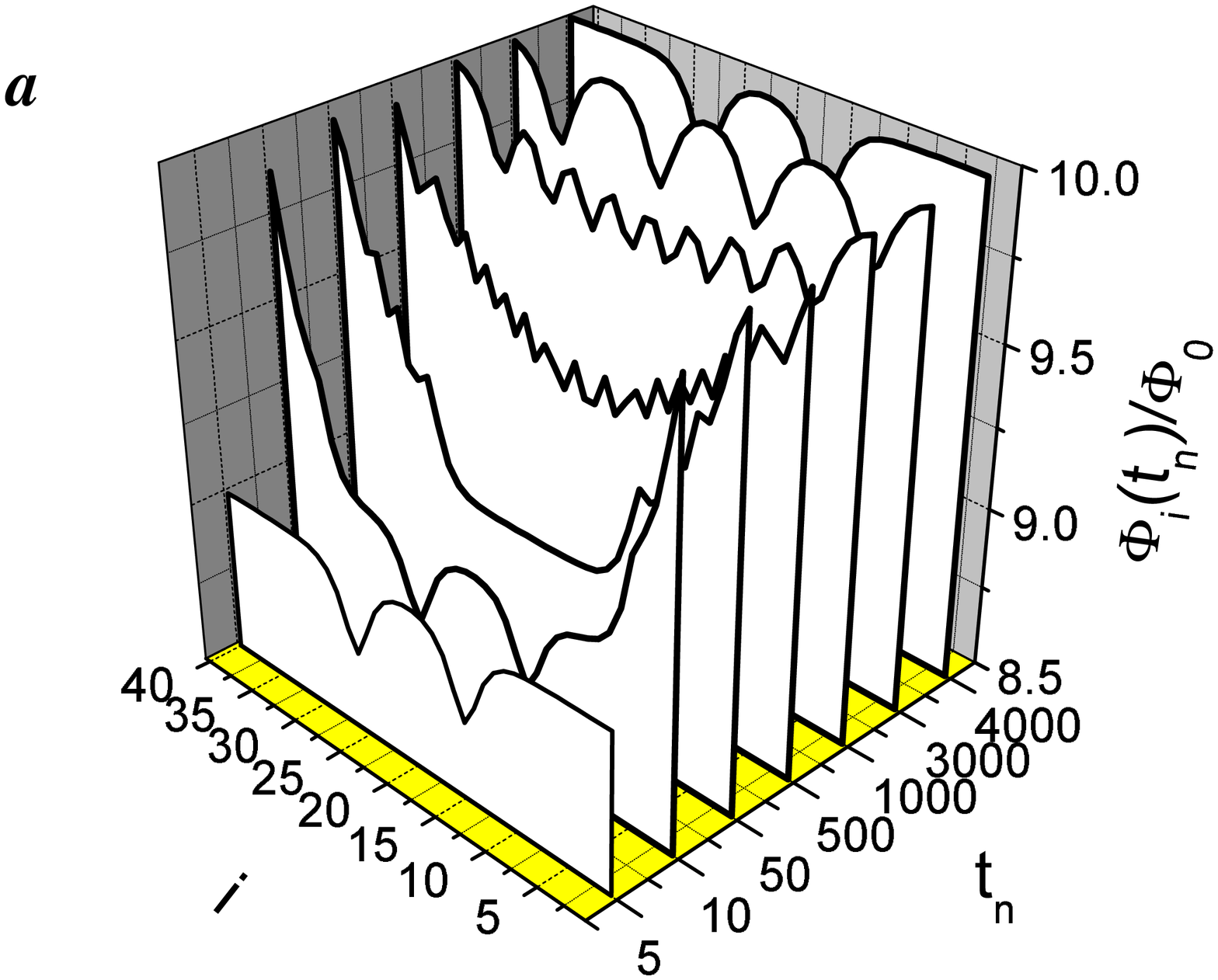}}
\centerline{\epsfxsize=8cm \epsfysize=8cm\epsfbox{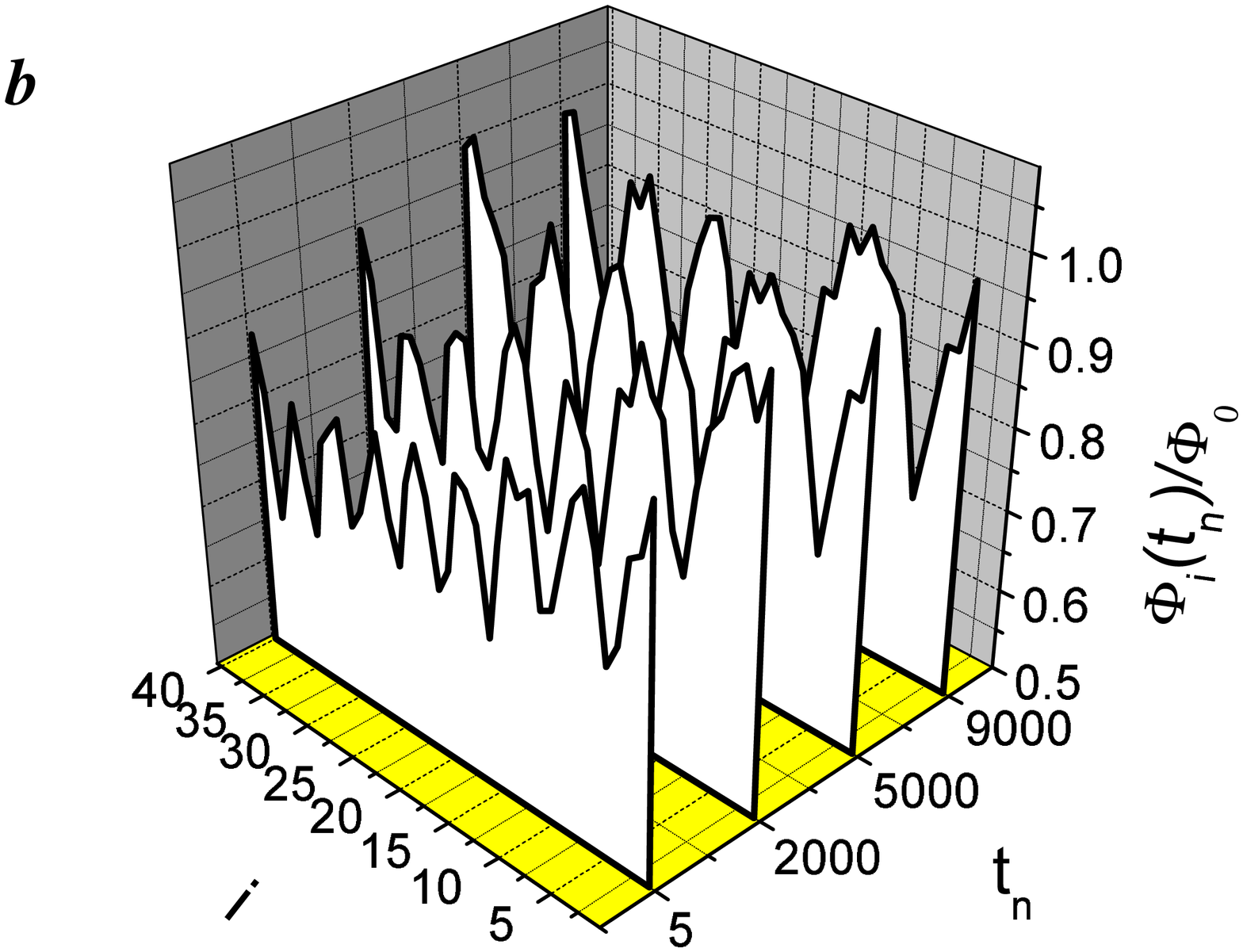}}
\caption{The  distribution of the magnetic flux in different moments
during a single avalanche. $V=0.6$, $N=40$. Time is measured in
arbitrary units. (a)~$\Delta J=0$; (b)~$\Delta J=0.1$.
\label{fig:fig101}}
\end{figure}

Fig. \ref{fig:fig101}  demonstrates the  process of penetration of
magnetic flux inside the SQUID and its redistribution between the
cells during the avalanche. The results presented in Fig.
\ref{fig:fig101}a correspond to a regular SQUID with $\Delta J=0$. The
figure shows the magnetic flux distributions corresponding to several
different moments in time during the avalanche. We see that at the
beginning of the avalanche the magnetic flux penetrates into the
boundary cells and only with time the penetration reaches  the central
part of the SQUID. The analogous results for a disordered system are
presented in Fig.~\ref{fig:fig101}b.  In this case, the  magnetic  flux
demonstrates the same dynamics as for a  regular system.  However, due
to the randomness the magnetic flux profile is much less regular than
in the previous case.

\begin{figure}
\centerline{\epsfxsize=6cm \epsfysize=9cm\epsfbox{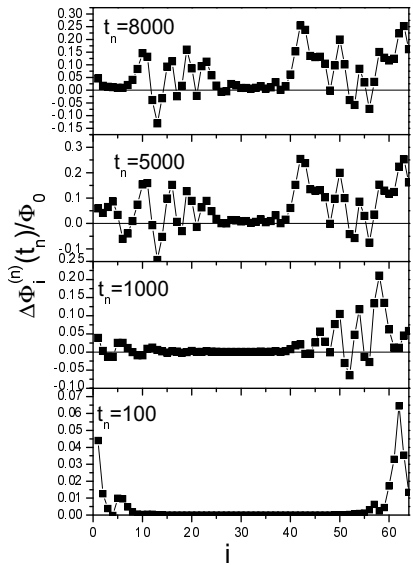}}
\vspace*{-0.7cm}
\caption{The magnetic flux differences $\Delta
\Phi_{i}^(n)(t_{n})/\Phi_{0}$ during a single avalanche. $V=0.6$, $N=65$
and $\Delta J=0.1$.
\label{fig:fig102}}
\end{figure}

If  $V\lesssim1$, each flux quantum may be considered as distributed
between several neighboring cells. In some cases, motion of such
extended vortices results in negative values of $\Delta\Phi$, as may
be seen in Fig.~\ref{fig:fig102}. This is true for both regular and
disordered SQUID's.

\vspace*{-0.2cm}

\section{The avalanche statistics in the critical state and
the self-organized criticality}

\begin{figure}
\centerline{\epsfxsize=6cm \epsfysize=6cm\epsfbox{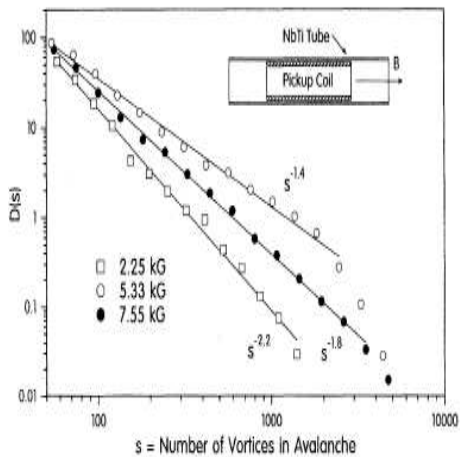}}
\caption{Distribution functions of the magnetic flux jumps obtained in
Ref. \onlinecite{Feld}.
\label{fig:Feld}}
\end{figure}

In experimental works on the critical state of superconductors, the
authors often points out  on the similarity of the system behavior  and
the phenomenon of self-organized criticality. Besides the
avalanche-like dynamics, it is a power-law distribution of magnetic
flux jumps. For instance, Fig. \ref{fig:Feld} shows the distribution
of avalanche sizes from Ref. \onlinecite{Feld}. In this work, the
 avalanche-like dynamics of magnetic flux in a  NbTi tube was studied.
Three distribution functions correspond to three fixed values of the
external magnetic field. The external magnetic field varies in the
interval of 30\,Oe centered at one of the fixed values with the rate of
5\,Oe/s.

\begin{figure}
\centerline{\epsfxsize=7cm\epsfysize=11cm\epsfbox{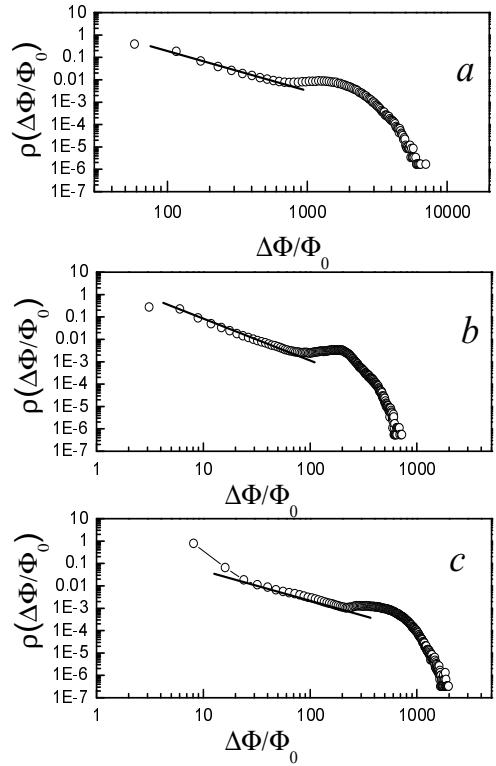}}
\caption{Probability densities for the magnetic flux jumps
$\rho(\Delta\Phi/\Phi_{0})$ for $\Delta J=0.5$.  (a)~$V=40$,
(b)~$V=1.2$, (c)~$V=0.6$. The straight lines represent a power law:
$(\Delta\Phi/\Phi_{0})^{\alpha}$ with $\alpha = -1.01, -1.26, -1.59$
 for (a), (b) and (c), respectively. \label{fig:fig5}}
\end{figure}

Fig. \ref{fig:fig5} demonstrates the results of our calculations for
the probability densities of jumps of the total magnetic flux for
multijunction SQUID for three different values of the parameter $V$.
 As was shown earlier,\cite{JLTP} the case of  $V=40$ for this degree
of disorder demonstrates the self-organized behavior. As may be seen
in Fig.~\ref{fig:fig5}, there are rather extended intervals of
power-law distributions for all considered values of $V$.

At the same time, from the point of view of classical  interpretation
of SOC  the magnitude of flux jumps is not a direct analog of the
avalanche size.\cite{Bak} However, just as the classical
self-organized critical state, the critical state in real
superconductors, as well as in our model, is self-reproducing and
consists of a large number of metastable states that transform to each
other by means of avalanches.  Thus, we can conclude that here we are
dealing with a more general type of a self-organized critical state,
in which the total magnetic flux plays a role of the main
characteristic of the system.

\section{Conclusions}

 Based on dynamical equations describing the simplest model of a
discrete superconductor (1D disordered multijunction SQUID), we present
a theoretical description of the avalanche-like dynamics of the
magnetic flux in ``hard" type-II superconductors.  Different values of
the SQUID-parameter $V\sim j_c a^3/\Phi_{0}$ are considered.   For
all values of $V$, including $V\lesssim1$, the  critical state in
the multijunction SQUID can be considered as a generalized type of a
self-organized critical state. In contrast to the classical definition of
SOC,\cite{Bak} the main characteristic of generalized critical state
is a size of magnetic flux jumps.  This
quantity demonstrates a power-law distribution if some degree of
disorder is introduced into the system. Our results are in qualitative
agreement with experiments.

\section {Acknowledgments}
{ This work is supported by the Russian Foundation for
Basic Research (project No.  05-02-17626), the
Scientific Council "Superconductivity",
the State programs "Quantum Macrophysics" and "Strong correlated
electrons in semiconductors, metals, superconductors and magnetic
materials".}


\begin{thebibliography}{99}


\bibitem{Bean} C.P. Bean, Phys. Rev. Lett. {\bf8}, 250 (1964).

\bibitem{DeGennes} DeGennes, P.G., {\em Superconductivity of Metals and
Alloys}, Benjamin, NY, 1966.

\bibitem{Alt} E. Altshuler and T.N. Johansen,
Rev. Mod. Phys. {\bf76}, 471 (2004)

\bibitem{Feld}S. Field, J. Witt, F.~Nori, X.~Ling, Phys. Rev. Lett.
{\bf74}, 1206 (1995).

\bibitem{Aeger} C.M. Aegerter, M.S. Welling, R.J.~Wijngaarden,
Europhys. Lett. {\bf65} 753 (2004).

\bibitem{Matizen}S.M. Ishikaev, E.V. Matizen, V.V.~Ryazanov,
V.A.~Oboznov, A.A.~Veretennikov, JETP Lett. {\bf72}, 39 (2000).

\bibitem{Bak}
P. Bak, C. Tang, K. Wiesenfeld, Phys. Rev. Lett. {\bf59}, 381 (1987).

\bibitem{Dhar} D. Dhar, Phys. Rev. Lett. {\bf64}, 1613 (1990).

 \bibitem{Ginzburg}
 S.L. Ginzburg,   JETP {\bf79}, 334 (1994).

 \bibitem{JETF} S.L. Ginzburg and N.E. Savitskaya,  JETP
 {\bf90}, 202 (2000).

\bibitem{JLTP} S.L. Ginzburg and N.E.~Savitskaya, Jornal of Low
Temperature Physics {\bf130}, N 3/4, 333 (2003).

\bibitem{Chen} D.-X. Chen, A. Sanchez, A.~Hernando,
Phys. Rev. B {\bf50}, 13735 (1994).

\bibitem{Wolf}
T. Wolf, A. Manjhofer, Phys.Rev. B. {\bf47}, 5383 (1993).

\bibitem{Chen1} D.-X. Chen,  A. Hernando,
Phys. Rev. B {\bf49}, 465 (1994).

\bibitem{Held}
G.A. Held, D.H. Solina, D.T. Keane, W.J.~Haag, P.M.~Horn,
G.~Grinstein, Phys. Rev. Lett. {\bf65}, 1120 (1990).

\end{thebibliography}
\end{document}